\documentclass[12pt]{article}
\usepackage{amssymb,amsmath,epsfig}

\begin{document}

\title{\bf Stability of Thin-Shell Wormholes in Nonlinear Electrodynamics}

\author{Muhammad SHARIF$^1$ \thanks{msharif.math@pu.edu.pk} and Muhammad AZAM$^{1,2}$
\thanks{azammath@gmail.com}\\
$^1$ Department of Mathematics, University of the Punjab,\\
Quaid-e-Azam Campus, Lahore-54590, Pakistan.\\
$^2$ Division of Science and Technology, University of Education,\\
Township Campus, Lahore-54590, Pakistan.}

\date{}

\maketitle
\begin{abstract}
In this paper, we construct thin-shell wormholes by applying the cut
and paste procedure to a regular charged black hole in nonlinear
electrodynamics field. We discuss different physical aspects of
wormholes such as, the possible equation of state to matter shell,
attractive or repulsive nature of wormhole and total amount of
exotic matter required. The thin-shell equation of motion with and
without cosmological constant is also investigated under linearized
perturbation. Finally, we explore the stability regions interpreted
by the parameter $\beta$ (speed of sound) and conclude that there
are realistic stability regions for some fixed values of parameters.
\end{abstract}

\section{Introduction}

Traversable and thin-shell wormholes physics has received great
attention due to many astrophysical phenomenon$^{{1-3})}$. The idea
of traversable wormhole (a solution of the field equations with two
asymptotically flat regions connected by a throat) was suggested by
Morris and Thorne$^{{4})}$. The physical evidence of wormholes has
always remained a great problem due to the existence of exotic
matter around the throat. However, it was proposed that the amount
of exotic matter can be minimized by choosing suitable geometry of
the wormhole$^{{5})}$. The class of thin-shell wormholes are
physically interesting, obtained from two manifolds by cut and paste
technique$^{{6,7})}$.

The stability of wormholes against perturbations is an interesting
issue. Stability of static wormholes has been investigated by using
specific equation of state (EoS) or by considering a linearized
radial perturbations around a static solution. In this context,
Visser$^{{6})}$, Poisson and Visser$^{{8})}$ carried out a
linearized analysis of the Schwarzschild thin-shell wormhole. Ishak
and Lake$^{{9})}$ studied stability of thin-shell wormholes and
found stable solutions. Armendariz-Picon$^{{10})}$ explored stable
solution of spherically symmetric configuration against linear
perturbations, whereas Shinkai and Hayward$^{{11})}$ investigated
unstable solution to this class for nonlinear perturbations.

It is found that charge$^{{12})}$ and positive cosmological
constant$^{{13})}$ increase the stability region in the analysis of
linearized stability of spherically symmetric thin-shell wormholes.
Thibeault et al.$^{{14})}$ studied stability of thin-shell wormhole
in Einstein-Maxwell theory with a Gauss-Bonnet term. Eiroa$^{{15})}$
explored stable and unstable static solutions of spherically
symmetric thin-shell wormholes supported by generalized Chaplygin
gas. The stability of thin-shell wormholes has been investigated
with regular charged black hole$^{{16})}$ and charged black hole in
generalized dilaton-axion gravity$^{{17})}$. We have studied the
problem of stability for spherical and cylindrical configurations at
Newtonian and post-Newtonian approximations and thin-shell wormholes
with Chaplygin gas$^{{18})}$.

Besides the lack of observational evidence, wormholes are considered
in the family of stars and black holes. For instance, wormhole
constructed from two FRW universes can be interpreted as a wormhole
with two dynamical stars$^{{19})}$. One can address properties of
wormholes and black holes in a unified way such as, black holes
being described by a null outer trapped surface and wormholes by a
timelike outer trapped surface (throat) where incoming null rays
start to diverge$^{{20})}$. Wormhole solutions have also been
studied in modified theories of gravity. Chodos and
Detweiler$^{{21})}$ and Clement$^{{22})}$ found solutions in higher
dimensions, whereas Nandi et. al$^{{23})}$ in Brans-Dicke theory and
Shen et. al$^{{24})}$ in Kaluza-Klein theory. Kar$^{{25})}$ and
Anchordoqui and Bergliaffa$^{{26})}$ found wormhole solutions in
Einstein-Gauss-Bonnet and brane world scenario respectively.

Born and Infeld$^{{27})}$ proposed a specific model based on a
principle of finiteness (to avoid physical quantities becoming
infinite) in nonlinear electrodynamics. After that,
Plebanski$^{{28})}$ presented some examples of nonlinear
electrodynamics Lagrangians and showed that the Born-Infeld theory
satisfies physically acceptable requirements. It is interesting to
note that the first exact regular black hole solution in general
relativity was found with nonlinear electrodynamics
source$^{{29,30})}$. Nonlinear Electrodynamics has been used widely
in many applications such as, these theories appear as effective
theories at different levels of string theory$^{{31})}$, explaining
the inflationary epoch and the late accelerated expansion of the
universe$^{{32,33})}$ and the effects produced by nonlinear
electrodynamics in spacetimes conformal to Bianchi
metrics$^{{34})}$. Also, homogenous and isotropic models have been
investigated with positive cosmological constant in this
scenario$^{{35})}$. It was shown that general relativity coupled to
nonlinear electrodynamics leads to regular magnetic black holes and
monopoles$^{{36})}$. The stability of regular electrically charged
structures$^{{37})}$ with regular de Sitter center has also been
explored$^{{38})}$. Baldovin et al.$^{{39})}$ have shown that an
effective wormhole geometry for an electromagnetic wave can appear
as a result of the nonlinear character of the field.

The objective of this paper is to construct spherically symmetric
thin-shell wormholes from regular charged black hole. To this end,
we use the Darmois-Israel thin-shell formalism to derive the
thin-shell equation of motion. We discuss various properties of
wormholes and investigate its stability regions under linearized
radial perturbation. The format of the paper is as follows. Section
\textbf{2} belongs to overview of the regular charged black hole.
Section \textbf{3} deals with the mathematical construction of
thin-shell wormhole. We discuss EoS relating pressure and density,
effect of gravitational field and calculate the total amount of
exotic matter in section \textbf{4}, when cosmological constant is
zero. Section \textbf{5} is devoted for the stability of wormhole.
In section \textbf{6}, we formulate thin-shell wormholes with
cosmological constant and find its stability. Finally, section
\textbf{7} discusses the results of the paper.

\section{Regular Charged Black Hole: An Overview}

The study on global regularity of black hole solution is quite
important in order to understand the final state of gravitational
collapse of initially regular configurations. None of the regular
black holes$^{{40-43})}$ (referred to Bardeen black holes) are exact
solutions to the Einstein field equations without any physically
reasonable source. It is well-known that electrovacuum
asymptotically flat metrics endowed with timelike and spacelike
symmetries do not allow for the existence of regular black hole
solutions. In order to derive regular black hole gravitational
nonlinear electromagnetic fields, one has to enlarge the class of
electrodynamics to nonlinear ones$^{{29})}$. These regular black
holes asymptotically behave as ordinary Reissner-Nordstr$\ddot{o}$m
black hole solution and the existence of these solutions does not
contradict with the singularity theorems$^{{44})}$. This motivates
us to construct a realistic thin-shell wormhole in nonlinear
electrodynamics and investigate its stability.

The action of gravitational field coupled to a nonlinear
electrodynamics field with cosmological constant can be defined
as$^{{45})}$
\begin{equation}\label{1}
S=\frac{1}{4\pi}\int{\sqrt{-g}}d^4x\left[\frac{1}{4}(R-2\Lambda)-\mathcal{L}(F)\right],
\end{equation}
where $R$ is the Ricci curvature of metric $g_{\alpha\beta}$,
$\Lambda$ is the cosmological constant and $\mathcal{L}(F)$ is a
gauge-invariant electromagnetic Lagrangian depending on a single
invariant $F=\frac{1}{4}F_{\alpha\beta}F^{\alpha\beta}$, where
$F_{\alpha\beta}= A_{\beta,\alpha}-A_{\alpha,\beta}$ is the
electromagnetic tensor. It is worth mentioning that nonlinear
electrodynamics source in the weak field limit becomes the Maxwell
field, i.e., $\mathcal{L}(F)=-\frac{F}{4\pi}$. Using the dual
transformation,
\begin{equation*}
\{g_{\alpha\beta}, ~F_{\alpha\beta}, ~F,
~\mathcal{L}(F)\}\leftrightarrow\{g_{\alpha\beta},
*P_{\alpha\beta}, -P, -\mathcal{H}(P)\}
\end{equation*}
where $*$ is a Hodge operator and
$P=\frac{1}{4}P_{\alpha\beta}P^{\alpha\beta}=(\mathcal{L}{_F})^2F$
with $P_{\alpha\beta}=\mathcal{L}{_F}F_{\alpha\beta}$ an
antisymmetric tensor and $\mathcal{L}_F=\frac{d\mathcal{L}}{dF}$,
the Hamiltonian $\mathcal{H}$ as a function of $P$ can be written as
$$d\mathcal{H}=(\mathcal{L}{_F})^{-1}d[(\mathcal{L}{_F})^2F]=\mathcal{H}_{P}dP.$$
Also, the Legendre transformation for Hamiltonian $\mathcal{H}$ and
Lagrangian $\mathcal{L}$ is$^{{46})}$
\begin{equation*}
\mathcal{H}=2F\mathcal{L}{_F}-\mathcal{L}, \quad
\mathcal{L}=2P\mathcal{H}{_P}-\mathcal{H}.
\end{equation*}

Einstein nonlinear electrodynamics field equations and Bianchi
identities in $F$ and $P$ are given, respectively, as follows
\begin{eqnarray}\label{a}
G_{\alpha\beta}=2[\mathcal{L}_{F}F^{\mu}_{\alpha}F_{\beta\mu}
-g_{\alpha\beta}(\mathcal{L}+\frac{1}{2}\Lambda)],\\\label{b}
\nabla_\alpha({\mathcal{L}_FF^{\alpha\mu}})=0,\quad
\nabla_\alpha{*{F^{\alpha\mu}}}=0,
\end{eqnarray}
\begin{eqnarray}\label{c}
G^\beta_\alpha=2[\mathcal{H}_{P}P_{\alpha\lambda}P^{\beta\lambda}
-\delta^\beta_\alpha(2P\mathcal{H}_P-\mathcal{H}+\frac{1}{2}\Lambda)],\\\label{d}
\nabla_\alpha{P^{\alpha\mu}}=0,\quad
\nabla_\alpha({*\mathcal{H}_P{P^{\alpha\mu}}})=0.
\end{eqnarray}
The specific function determining the nonlinear electromagnetic
source is
\begin{equation*}
\mathcal{H}(P)=P\frac{(1-3\sqrt{-2Q^2P})}{(1+\sqrt{-2Q^2P})^3}-
\frac{3}{2Q^2s}\left(\frac{\sqrt{-2Q^2P}}{(1+\sqrt{-2Q^2P})}\right)^\frac{5}{2}.
\end{equation*}
The corresponding nonlinear electromagnetic Lagrangian for the
action (\ref{1}) in $F$ and $P$ frameworks becomes$^{{45})}$
\begin{equation*}
\mathcal{L}(F)=F\frac{(1-3\sqrt{2Q^2F})}{(1+\sqrt{2Q^2F})^3}
+\frac{3}{2Q^2s}\left(\frac{\sqrt{2Q^2F}}{1+\sqrt{2Q^2F}}\right)^\frac{5}{2},
\end{equation*}
\begin{equation*}
\mathcal{L}(P)=P\frac{(1-8\sqrt{-2Q^2P}-6Q^2P)}{(1+\sqrt{-2Q^2P})^4}-
\frac{3}{4Q^2s}\frac{(-2Q^2P)^{\frac{5}{4}}(3-2\sqrt{-2Q^2P})}
{\left(1+\sqrt{-2Q^2P}\right)^\frac{7}{2}},
\end{equation*}
where $s=\frac{\mid{Q}\mid}{2M},~M$ and $Q$ are free parameters
which would correspond to mass and charge respectively.

Next, we obtain a compatible solution for the $\mathcal{L}(F)$. For
this purpose, we consider a static spherically symmetric
configuration
\begin{equation}\label{e}
\mathfrak{g}=-\left(1-\frac{2m(r)}{r}\right)dt^2+\left(1-\frac{2m(r)}{r}\right)^{-1}dr^2
+r^2(d\theta^{2}+\sin^2{\theta}d\phi^2).
\end{equation}
Integration of the first of Eq.(\ref{d}) with the ansatz for the
antisymmetric field
$P_{\alpha\beta}=2\delta^t_{[\alpha}\delta^r_{\beta]}D(r)$ yields
$P=-\frac{Q^2}{2r^4}$, where we have chosen $Q$ as an integration
constant which plays the role of charge in the bulk solution. The
zero-zero component of Eq.(\ref{a}) yields
\begin{equation}\label{f}
m'(r)=r^2\mathcal{L}(F)+\frac{\Lambda{r^2}}{2}.
\end{equation}
Using $F=-P=\frac{Q^2}{2r^4}$ as well as $\mathcal{L}(F)$ in this
equation and then integrating, we have
\begin{equation}\label{g}
m(r)=M-3MQ^2\int^{\infty}_r\frac{Q^2}{(y^2+Q^2)^\frac{5}{2}}dy
-\frac{Q^2}{2}\int^{\infty}_r\frac{(y^4-3Q^2y^2)}{(y^2+Q^2)^3}dy+\frac{\Lambda{r^3}}{6},
\end{equation}
where we have used $M=\lim_{r\rightarrow{\infty}}m(r)$. After
integration, it follows that
\begin{equation}\label{h}
m(r)=\frac{Mr^3}{(r^2+Q^2)^\frac{3}{2}}
-\frac{Q^2r^3}{2(r^2+Q^2)^2}+\frac{\Lambda{r^3}}{6}.
\end{equation}
Inserting $m(r)$ in Eq.(\ref{e}), we find a solution compatible with
the Lagrangian $\mathcal{L}(F)$ as follows
\begin{equation}\label{1a}
ds^2=-f(r)dt^{2}+f^{-1}(r)dr^{2}+r^2(d\theta^{2}+\sin^2{\theta}d\phi^2),
\end{equation}
where
$f(r)=1-\frac{2Mr^2}{(r^2+Q^2)^{\frac{3}{2}}}+\frac{Q^2r^2}{(r^2+Q^2)^2}
-\frac{\Lambda{r^2}}{3}$ is a positive function for the given
radius.

Now we show that this solution satisfies the weak energy condition.
The stress-energy tensor from Eq.(\ref{a}) can be written as
\begin{equation}\label{1b}
T_{\alpha\beta}=F_{\alpha\gamma}F^{\gamma}_\beta{\mathcal{L}_F}-g_{\alpha\beta}\mathcal{L}(F),
\end{equation}
where the electromagnetic field tensor defined as
\begin{equation*}
F_{\alpha\beta}=E(r)(\delta^t_\alpha\delta^r_\beta-\delta^r_\alpha\delta^t_\beta).
\end{equation*}
The invariant $F$ can then be written as
\begin{equation}\label{1c}
2F=-E^2(r),
\end{equation}
which shows that the electric  field can be expressed in terms of
the invariant $F$.  To deal with the physically reasonable theories,
the fulfillment of the weak energy condition is sufficient for the
corresponding energy-momentum tensor. For this, one requires
$T_{\alpha\beta}u^\alpha{u^\beta}\geq0$ and
$Q_{\alpha}Q^\alpha\leq0$, where $Q^\alpha=T^\alpha_\beta{u^\beta}$
for timelike vector
$u^\alpha=\frac{\delta^\alpha_t}{\sqrt{-g_{tt}}}$. The first
inequality implies
\begin{equation*}
(\mathcal{L}+E^2\mathcal{L}_{F})\geq0,
\end{equation*}
which becomes $\mathcal{L}\geq{E}\mathcal{L}_{E}$ using
Eq.(\ref{1c}). The second inequality becomes
\begin{equation*}
-(\mathcal{L}+E^2\mathcal{L}_{F})^2\leq0,
\end{equation*}
which shows that norm of the energy flux $Q^\alpha$ occurs to be
less than zero.

\section{Thin-Shell Wormhole}

For the mathematical construction of thin-shell wormhole, we cut two
identical $4$D copies of the regular charged black hole as
\begin{equation}\label{2}
\mathcal{M}^{\pm}=\{x^{\gamma}=(t,r,\theta,\phi)\mid{r}\geq{a}\},
\end{equation}
where $`a$' is the throat radius. We paste these geometries at the
timelike hypersurface $\Sigma=\Sigma^\pm=\{r-a=0\}$ to get a new
geodesically complete manifold
$\mathcal{M}=\mathcal{M}^{+}\cup{\mathcal{M}^{-}}$ with matter shell
at $r=a$. The resulting manifold describes a wormhole with two
regions connected by a throat radius placed in the joining shell
satisfying the flare-out condition$^{{47})}$.

We follow the standard Darmois-Israel formalism$^{{48,49})}$ to
understand the dynamics of the wormhole. The induced metric with
throat radius as a function of proper time on $\Sigma$ is defined as
\begin{equation}\label{3}
ds^2=-d\tau^2+a^2(\tau)(d\theta^2+\sin^2{\theta}d\phi^2).
\end{equation}
The two sides of the shell are matched through the extrinsic
curvature defined on $\Sigma$ as
\begin{equation}\label{4}
K^{\pm}_{ij}=-n^{\pm}_{\gamma}(\frac{{\partial}^2x^{\gamma}_{\pm}}
{{\partial}{\xi}^i{\partial}{\xi}^j}+{\Gamma}^{\gamma}_{{\mu}{\nu}}
\frac{{{\partial}x^{\mu}_{\pm}}{{\partial}x^{\nu}_{\pm}}}
{{\partial}{\xi}^i{\partial}{\xi}^j}),\quad(i, j=0,2,3),
\end{equation}
where $\xi^i=(\tau,\theta,\phi)$ are the coordinates with $\tau$ as
the proper time on the shell and $n^{\pm}_{\gamma}$ are the unit
normals satisfying the relation $n^{\gamma}n_{\gamma}=1$ to $\Sigma$
given by
\begin{equation}\label{5}
n^{\pm}_{\gamma}=\left(-\dot{a},\frac{\sqrt{f(a)+\dot{a}^2}}{f(a)},0,0\right).
\end{equation}
Using the orthonormal basis
$\{e_{\hat{\tau}}=e_{\tau},~e_{\hat{\theta}}=a^{-1}e_{\theta},~
e_{\hat{\phi}}=(a\sin{\theta})^{-1}e_{\phi}\}$, the non-vanishing
components of the extrinsic curvature turns out to be
\begin{equation}\label{6}
K^{\pm}_{\hat\tau\hat\tau}=\mp\frac{f'(a)+2\ddot{a}}{2\sqrt{f(a)+\dot{a}^2}},
\quad
K^{\pm}_{\hat\theta\hat\theta}=K^{\pm}_{\hat\phi\hat\phi}={\pm}\frac{1}{a}\sqrt{f(a)+\dot{a}^2}.
\end{equation}
Here dot and prime denote derivatives with respect to $\tau$ and $a$
respectively.

The surface stress-energy tensor
$S_{\hat{i}\hat{j}}=diag(\sigma,p_{\hat\theta},p_{\hat\phi})$
provides surface energy density $\sigma$ and surface tensions
$p_{\hat\theta},p_{\hat\phi}$ of the shell. The Lanczos equations
are defined on the shell as
\begin{equation}\label{7}
S_{\hat{i}\hat{j}}=\frac{1}{8\pi}\left\{g_{\hat{i}\hat{j}}K-[K_{\hat{i}\hat{j}}]\right\},
\end{equation}
where
$[K_{\hat{i}\hat{j}}]=K^{+}_{\hat{i}\hat{j}}-K^{-}_{\hat{i}\hat{j}}$
and $K=tr[K_{\hat{i}\hat{j}}]=[K^{\hat{i}}_{\hat{i}}]$. Using
Eqs.(\ref{6}) and (\ref{7}), the energy density $\sigma$ and
transverse pressure $p$ to shell are obtained as
\begin{eqnarray}\label{8}
\sigma&=&-\frac{1}{2\pi{a}}\sqrt{f(a)+\dot{a}^2},\\\label{9}
p&=&p_{\hat\theta}=p_{\hat\phi}=\frac{1}{8\pi{a}}\frac{2a\ddot{a}+2\dot{a}^2
+2f(a)+af'(a)}{\sqrt{f(a)+\dot{a}^2}}.
\end{eqnarray}
\begin{figure}
\centering \epsfig{file=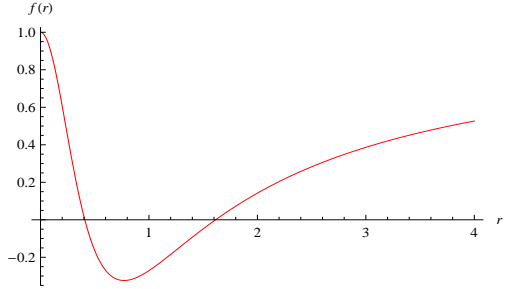,width=.48\linewidth, height=1.5in}
\caption{(Color online) $f(r)$ cuts the $r-$axis at $r_{-}$ and
$r_{+}$ (Event horizons) for $Q=0.5$ and $M=1$ in the absence of
$\Lambda$.}
\end{figure}

\section{Thin-Shell Wormhole Without $\Lambda$}

Here we consider the case, when $\Lambda=0$ for the given metric. It
is shown in Fig. \textbf{1} that there exist two event horizons for
such a black hole, whenever $|{Q}|\leq{0.6M}$$^{{29})}$. We explore
EoS parameter. The corresponding energy density and surface
potential from Eqs.(\ref{8}) and (\ref{9}) for static configuration
of radius $a$ ($\dot{a}=\ddot{a}=0$) become
\begin{eqnarray}\label{10}
\sigma&=&-\frac{1}{2\pi{a}}\left[1-\frac{2Ma^2}{(a^2+Q^2)^{\frac{3}{2}}}
+\frac{Q^2a^2}{(a^2+Q^2)^2}\right]^\frac{1}{2},\\\label{11}
p&=&\frac{\left[1-\frac{4Ma^2}{(a^2+Q^2)^{\frac{3}{2}}}
+\frac{3Ma^4}{(a^2+Q^2)^{\frac{5}{2}}}
+\frac{2Q^2a^2}{(a^2+Q^2)^2}-\frac{2Q^2a^4}{(a^2+Q^2)^3}\right]}
{4\pi{a}\left[1-\frac{2Ma^2}{(a^2+Q^2)^{\frac{3}{2}}}
+\frac{Q^2a^2}{(a^2+Q^2)^2}\right]^\frac{1}{2}}.
\end{eqnarray}
\begin{figure}
\centering \epsfig{file=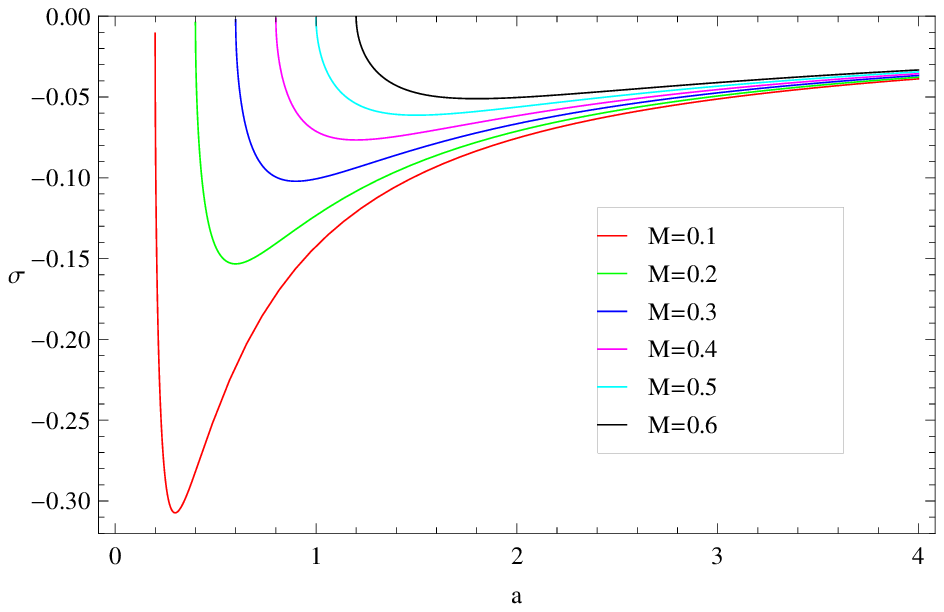,width=.48\linewidth, height=1.5in}
\epsfig{file=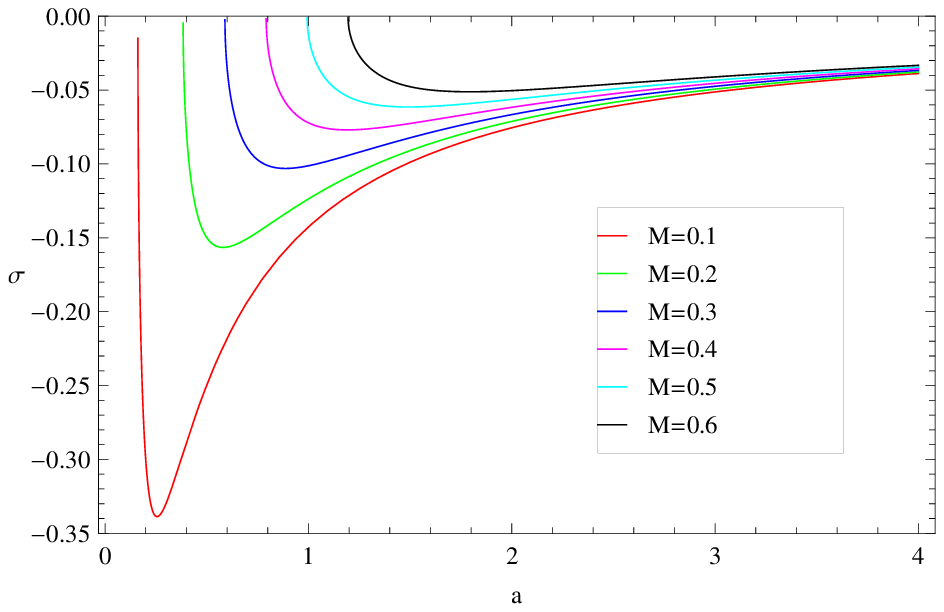,width=.48\linewidth, height=1.5in}
\caption{(Color online) Plots of $\sigma$ versus $a$ with fixed
values of $Q=0.01$ and $Q=0.05$.}
\end{figure}
\begin{figure}
\centering \epsfig{file=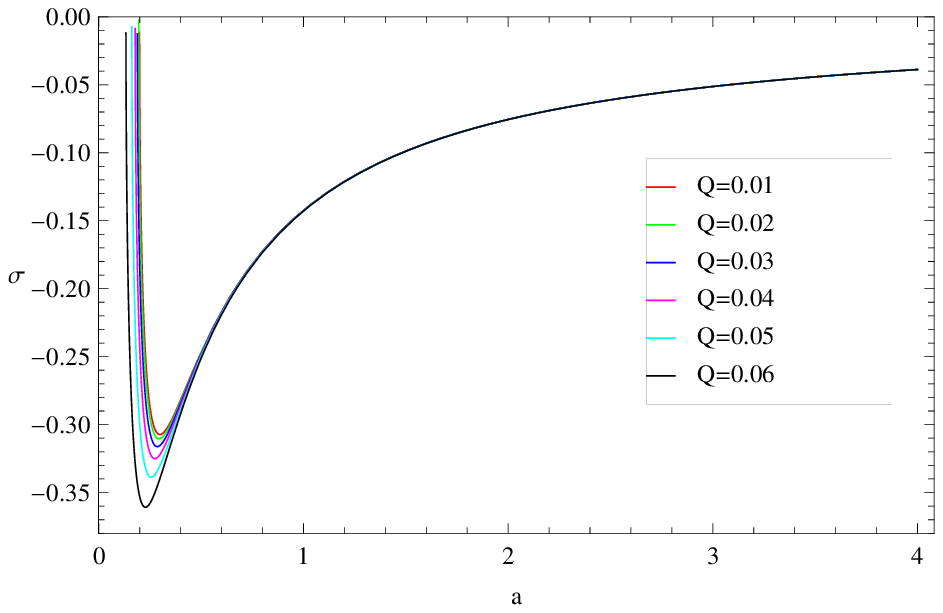,width=.48\linewidth, height=1.5in}
\epsfig{file=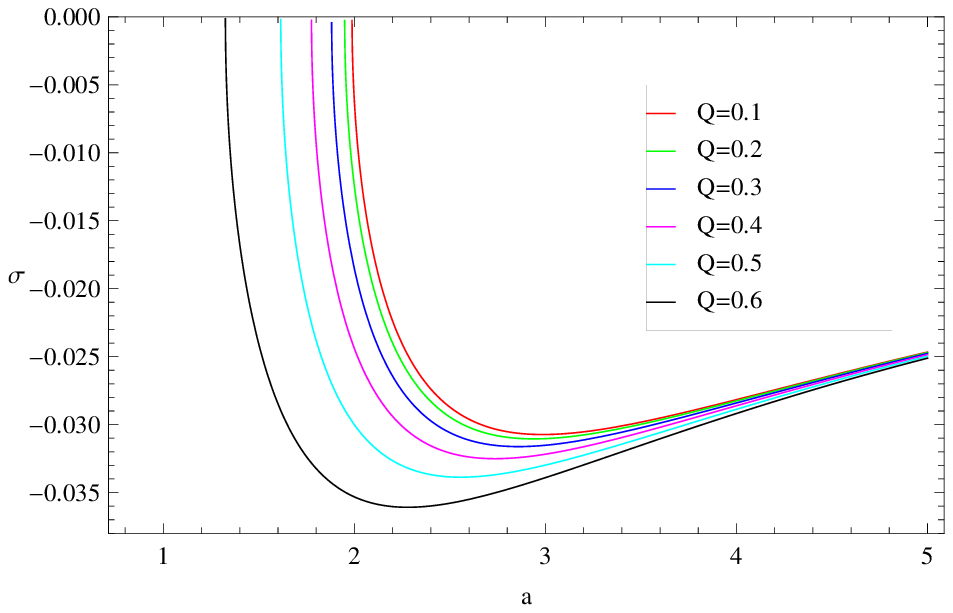,width=.48\linewidth, height=1.5in}
\caption{(Color online) Plots of $\sigma$ versus $a$ with fixed
values of $M=0.1$ and $M=1$.}
\end{figure}
\begin{figure}
\centering \epsfig{file=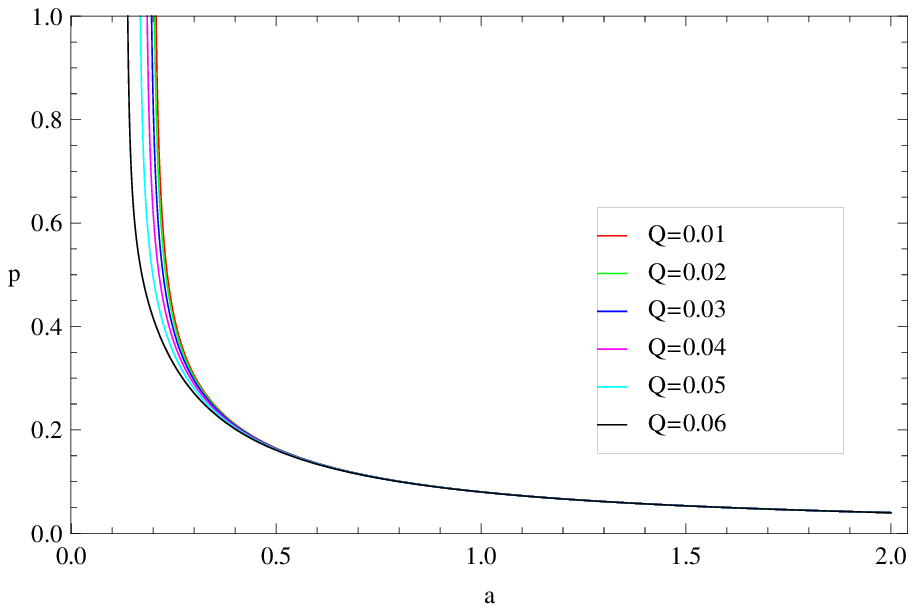,width=.48\linewidth, height=1.5in}
\epsfig{file=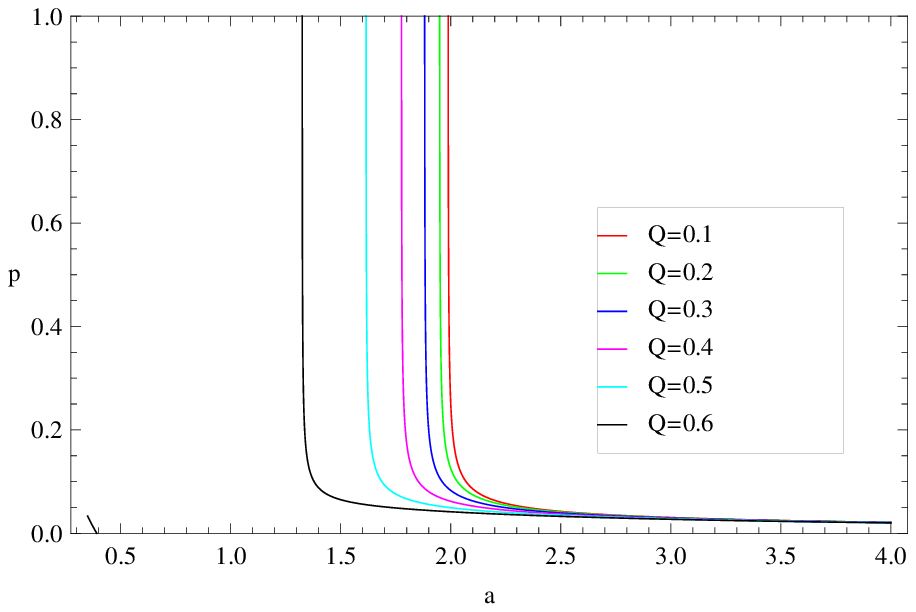,width=.48\linewidth,
height=1.5in}\caption{(Color online) Plots of $p$ versus $a$ for
fixed values of $M=0.1$ and $M=1$.}
\end{figure}
In static wormholes, the throat is defined as minimal area surface
satisfying the flare-out condition. To fulfill this condition,
wormholes must be threaded by the exotic matter (violating the null
energy condition (NEC), i.e., $\sigma+p<0$)$^{{50,51})}$. We see
from Figs. \textbf{2-5} that energy density is negative and pressure
is a decreasing function of throat radius for some fixed values of
charge and mass. This shows that matter on the shell (violating both
NEC and weak energy condition (WEC)) implies the presence of exotic
matter. For the explanation of this matter, we consider an EoS on
the shell as
\begin{figure}
\centering \epsfig{file=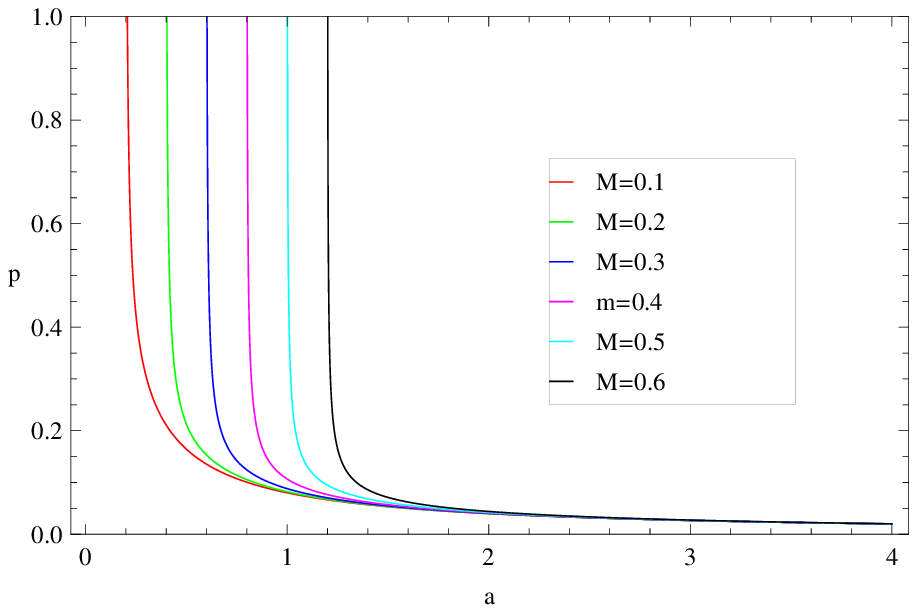,width=.48\linewidth, height=1.5in}
\epsfig{file=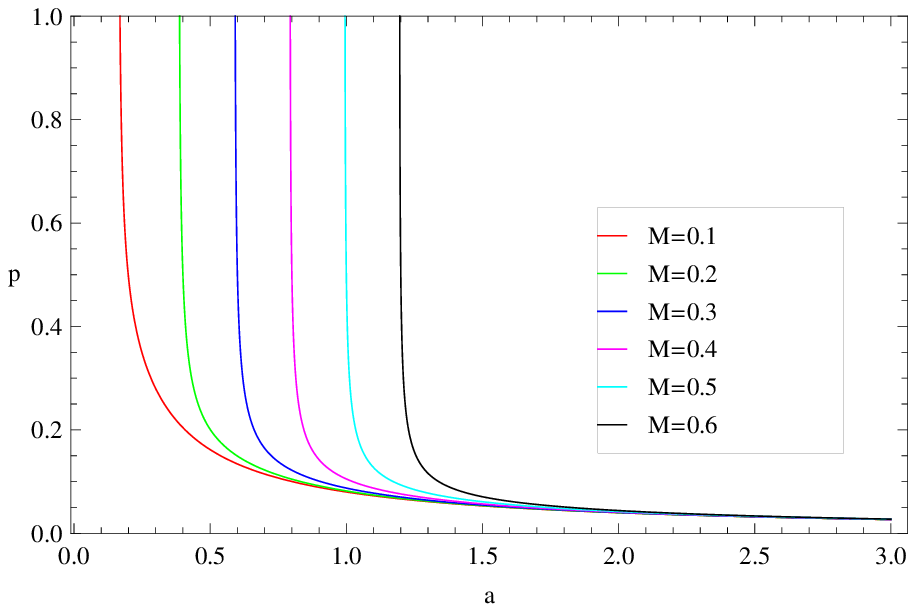,width=.48\linewidth,
height=1.5in}\caption{(Color online) Plots of $p$ versus $a$ for
fixed value of $Q=0.01$ and $Q=0.05$.}
\end{figure}
\begin{equation}\label{12}
p=\omega\sigma,
\end{equation}
where $\omega$ is the EoS parameter. Using Eqs.(\ref{10}) and
(\ref{11}), we have
\begin{eqnarray}\label{13}
\frac{p}{\sigma}=\omega=-\frac{1}{2}-\frac{\left[-\frac{2Ma^2}{(a^2+Q^2)^{\frac{3}{2}}}
+\frac{3Ma^4}{(a^2+Q^2)^{\frac{5}{2}}}
+\frac{Q^2a^2}{(a^2+Q^2)^2}-\frac{2Q^2a^4}{(a^2+Q^2)^3}\right]}
{2\left[1-\frac{2Ma^2}{(a^2+Q^2)^{\frac{3}{2}}}
+\frac{Q^2a^2}{(a^2+Q^2)^2}\right]}.
\end{eqnarray}

We see that $\omega\rightarrow{-\frac{1}{2}}$, if the location of
the wormhole throat is large enough, i.e., $a\rightarrow\infty$.
This shows that the distribution of matter in the shell is of dark
energy type. We obtain dust shell, i.e., $p\rightarrow{0}$ when
$a\rightarrow{a_0}$ (the point where the curve cuts the $a$-axis
shown in the right graph of Fig. \textbf{6}) for fixed values of
$M=1$ and $Q=0.5$. However, $a_0<r_h$ (in our case), the dust shell
can never be found. Also, the Casimir effect with massless field is
of traceless type, so the dust shell can be looked into by taking
trace of the surface stress-energy tensor, i.e., $S^i_j=0$ implies
that $-\sigma+2p=0$, which yields
\begin{figure} \centering
\epsfig{file=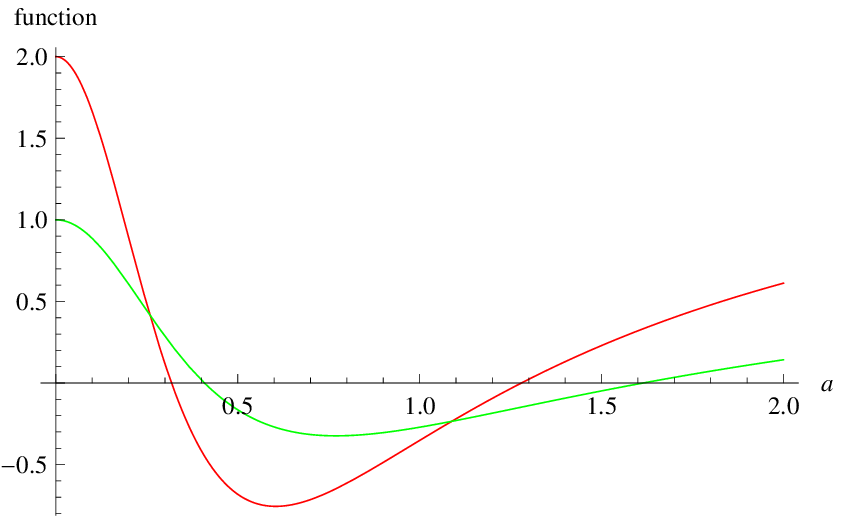,width=.48\linewidth, height=1.8in}
\epsfig{file=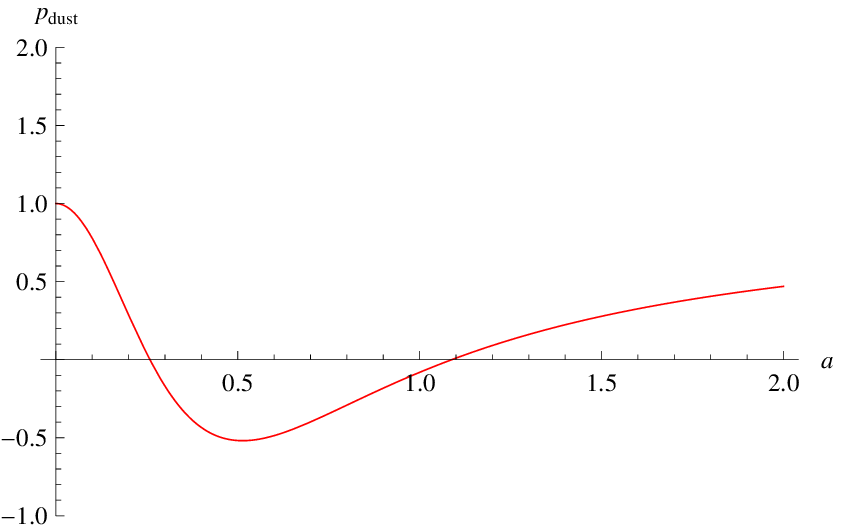,width=.48\linewidth,
height=1.8in}\caption{(Color online) The left graph shows the
behavior of the function $f(a)$ (green) and $g(a)$ (red) for fixed
values of $Q=0.5$ and $M=1$.}
\end{figure}
\begin{equation}\label{14}
g(a)=2-\frac{6Ma^2}{(a^2+Q^2)^\frac{3}{2}}+\frac{3Ma^4}{(a^2+Q^2)^\frac{5}{2}}+
\frac{3Q^2a^2}{(a^2+Q^2)^{2}}-\frac{2Q^2a^4}{(a^2+Q^2)^{3}}=0.
\end{equation}
The left graph of Fig. \textbf{6} shows that the red curve cuts the
$a$-axis before the green curve. This shows that throat of the
wormhole satisfying the above equation lies inside the event horizon
of the black hole, hence not possible in a wormhole configuration.
Thus, a dust shell cannot be found.

\subsubsection*{The Gravitational Field}

The gravitational field of wormhole depends upon the observer's four
acceleration. The nature of wormhole is attractive or repulsive
according to $a^r>0$ or $a^r<0$, respectively. The observer's four
acceleration is defined as
\begin{equation}\label{15}
a^{\mu}=v^{\mu}_{;\nu}v^{\nu},
\end{equation}
where
\begin{equation*}\label{16}
v^{\nu}=\frac{dx^{\nu}}{d{\tau}}=\left(\frac{1}{\sqrt{f(r)}},0,0,0\right).
\end{equation*}
The only non-vanishing component of the four acceleration is
\begin{equation}\label{17}
a^{r}=\Gamma^{r}_{tt}\left(\frac{dt}{d{\tau}}\right)^2=\frac{1}{r}\beta(r),
\end{equation}
where
\begin{eqnarray}\label{18}
\beta(r)=\left[-\frac{2\frac{M}{r}}{(1+(\frac{Q}{r})^2)^{\frac{3}{2}}}
+\frac{3\frac{M}{r}}{(1+(\frac{Q}{r})^2)^{\frac{5}{2}}}
+\frac{(\frac{Q}{r})^2}{(1+(\frac{Q}{r})^2)^2}-\frac{2(\frac{Q}{r})^2}{(1+(\frac{Q}{r})^2)^3}\right].
\end{eqnarray}
Also, the the geodesic equation for a radially moving test particle
initially at rest can be defined as
\begin{equation}\label{19}
a^{r}=-\frac{d^2r}{d\tau^{2}}=-\Gamma^{r}_{tt}\left(\frac{dt}{d{\tau}}\right)^2.
\end{equation}
Figure \textbf{7} shows that the curve cuts at the point
$x=\frac{Q}{r}$ for a fixed value of $\frac{M}{r}=1$. Hence, the
wormhole will be attractive for $x>\frac{Q}{r}$ and repulsive for
$x<\frac{Q}{r}$. We obtain a geodesic equation for $x=\frac{Q}{r}$,
i.e., $a^r=0$.
\begin{figure} \centering
\epsfig{file=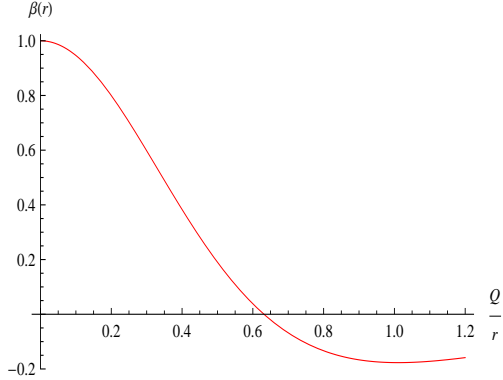,width=.48\linewidth, height=2in}
\caption{(Color online) Plot of $\beta(r)$ versus $\frac{Q}{r}$: $x$
is the point where the function $\beta(r)$ cuts the
$\frac{Q}{r}-$axis.}.
\end{figure}

\subsubsection*{The Total Amount of Exotic Matter}

It is found that the total amount of exotic matter required to
support wormholes can be made as small as possible by choosing
appropriate geometry of the wormhole$^{{5})}$. Here, we show that
the total amount of exotic matter can be reduced by either
increasing the mass or decreasing the charge of the black hole. We
consider the integral$^{{52})}$
\begin{equation}\label{20}
\Omega_\alpha=\int\left(\rho+p\right)\sqrt{-g}d^3x.
\end{equation}
Choosing $R=r-a$ as the radial coordinate, we have
\begin{equation}\label{21}
\Omega_\alpha=\int^{2\pi}_0\int^{\pi}_0\int^{\infty}_{-\infty}
\left(\rho+p\right)\sqrt{-g}dRd{\theta}d{\phi}.
\end{equation}
\begin{figure}
\centering \epsfig{file=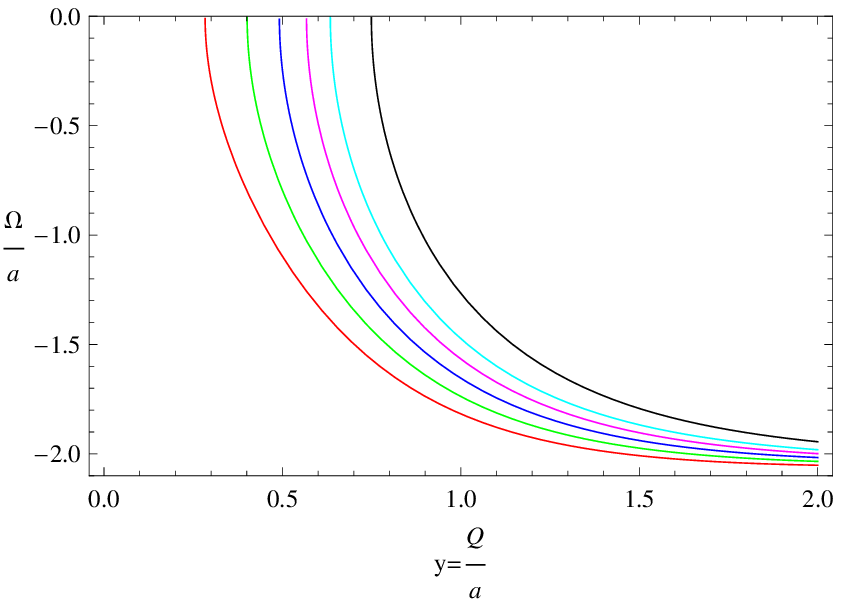,width=.48\linewidth, height=2in}
\epsfig{file=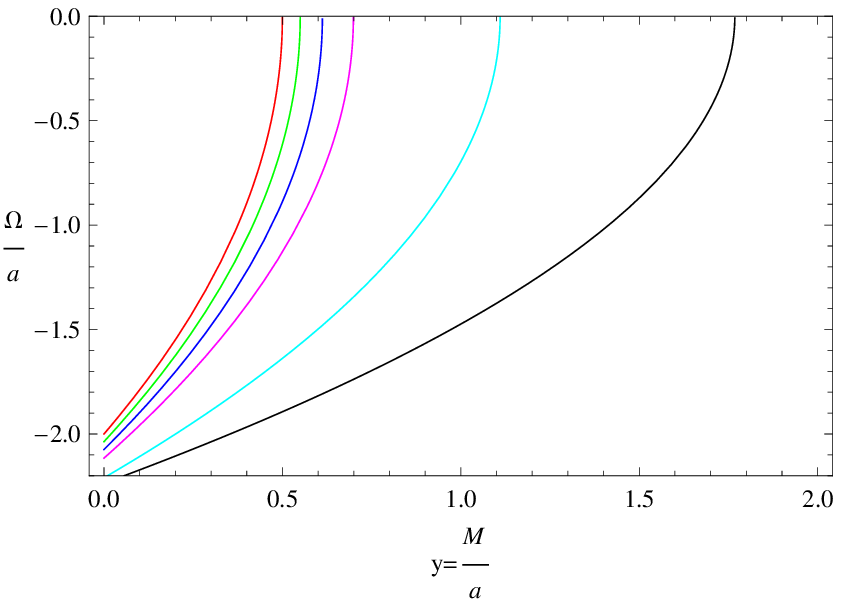,width=.48\linewidth,
height=2in}\caption{(Color online) Variation in the total amount of
exotic matter for $\frac{M}{a}=0.6, ~0.7, ~0.8, ~0.9, ~1, ~1.2$ and
$\frac{Q}{a}=0, ~0.2, ~0.3, ~0.4, ~0.7, ~1$.}.
\end{figure}
The infinitely thin shell does not exert any radial pressure,
therefore, using $\rho=\delta(R)\sigma(a)$, it follows that
\begin{eqnarray}\nonumber
\Omega_{\alpha}&=&\int^{2\pi}_0\int^{\pi}_0\sigma\sqrt{-g}|_{r=a}
d{\theta}d{\phi}=4\pi{a^2}\sigma(a)\\\label{22}
&=&-2a\left[1-\frac{2Ma^2}{(a^2+Q^2)^{\frac{3}{2}}}
+\frac{Q^2a^2}{(a^2+Q^2)^2}\right]^{\frac{1}{2}}.
\end{eqnarray}

The matter violating NEC can be reduced by choosing $a$ closer to
$r_h$. When $a\rightarrow{r_h}$, the wormhole is closer to a black
hole as the incoming microwave background radiation would become
blueshifted at a high temperature$^{{53})}$. It is observed from
Eq.(\ref{22}) that for $a\gg{r_h},~\Omega_{\alpha}$ will depend
linearly upon $a$, i.e., $\Omega_{\alpha}\approx{-2a}$. Figure
\textbf{8} shows the variation of exotic matter corresponding to
mass and charge of the black hole. Thus, the total exotic matter can
be reduced by either increasing mass or decreasing charge of the
black hole.

\section{Stability Analysis}

In this section, we obtain the stability regions of the static
configuration of wormhole under a small perturbation around the
static solution $a=a_0$. For this purpose, we consider thin shell
equation of motion (by re-arranging Eq.(\ref{8}))
\begin{equation}\label{23}
\dot{a}^2+V(a)=0,
\end{equation}
where the potential $V(a)$ is given as
\begin{equation}\label{24}
V(a)=f(a)-\left[2\pi{a}{\sigma(a)}\right]^2.
\end{equation}
For the linearized stability criteria, we use Taylor expansion to
$V(a)$ upto second order around $a_0$
\begin{eqnarray}\label{25}
V(a)=V(a_0)+V'(a_0)(a-a_0)+\frac{1}{2}V''(a_0)(a-a_0)^2+O[(a-a_0)^3].
\end{eqnarray}
The change in the internal energy plus the work done by the internal
forces of the throat is equal to the flux term known as energy
conservation equation, which can be written with the help of
Eqs.(\ref{8}) and (\ref{9}) as
\begin{eqnarray}\label{26}
\frac{d(\sigma\mathcal{A})}{d\tau}+p\frac{d\mathcal{A}}{d\tau}=\left\{[2a]^2-4a\right\}
\frac{\dot{a}\sqrt{f(a)+\dot{a}^2}}{4a},
\end{eqnarray}
where $\mathcal{A}=4\pi{a^2}$ is the area of the wormhole throat.
Using ${\sigma}'=\frac{\dot{\sigma}}{\dot{a}}$ in the above
equation, we have
\begin{equation}\label{27}
a^2{\sigma}'+2a(\sigma+p)+\left\{[2a]^2-4a\right\}\frac{\sigma}{4a}=0.
\end{equation}
Inserting $(a\sigma)'=-(\sigma+p)$, the first derivative of
Eq.(\ref{24}) can be written as
\begin{equation}\label{28}
V'(a)=f'(a)+8{\pi}^2a\sigma(\sigma+p).
\end{equation}

Here, we define a parameter $\beta$ (velocity of sound), which
should be less than one for a realistic model$^{{54,55})}$
\begin{equation}\label{29}
\beta^2(\sigma)=\frac{\partial{p}}{\partial{\sigma}}|_\sigma.
\end{equation}
Notice that
\begin{figure}
\centering \epsfig{file=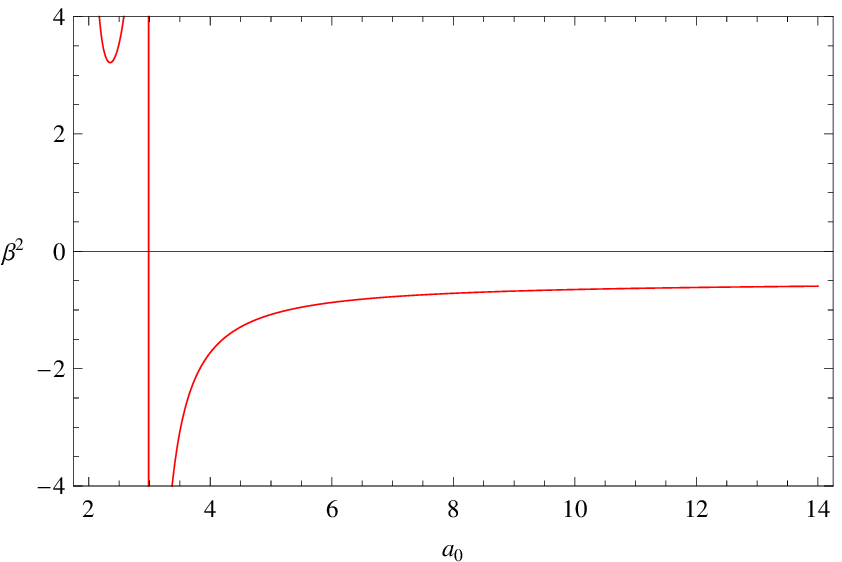,width=.30\linewidth, height=2in}
\epsfig{file=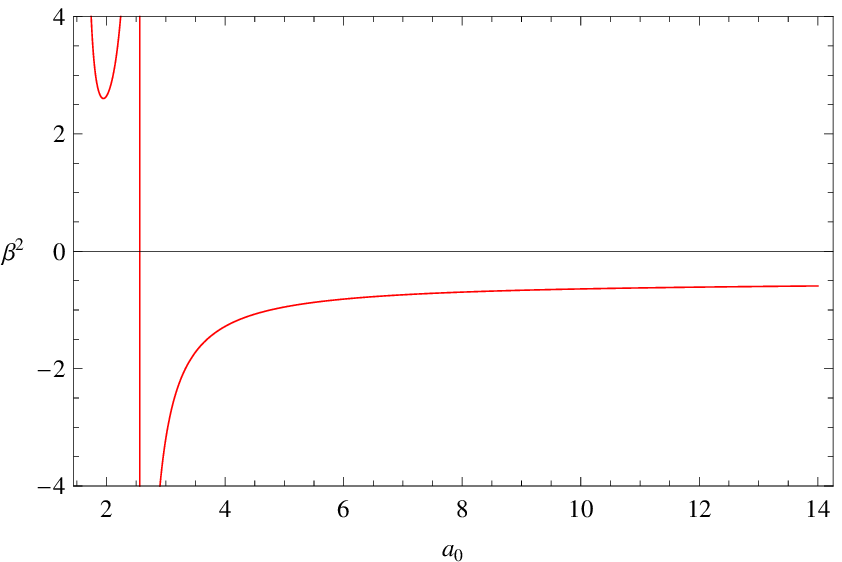,width=.30\linewidth, height=2in}
\epsfig{file=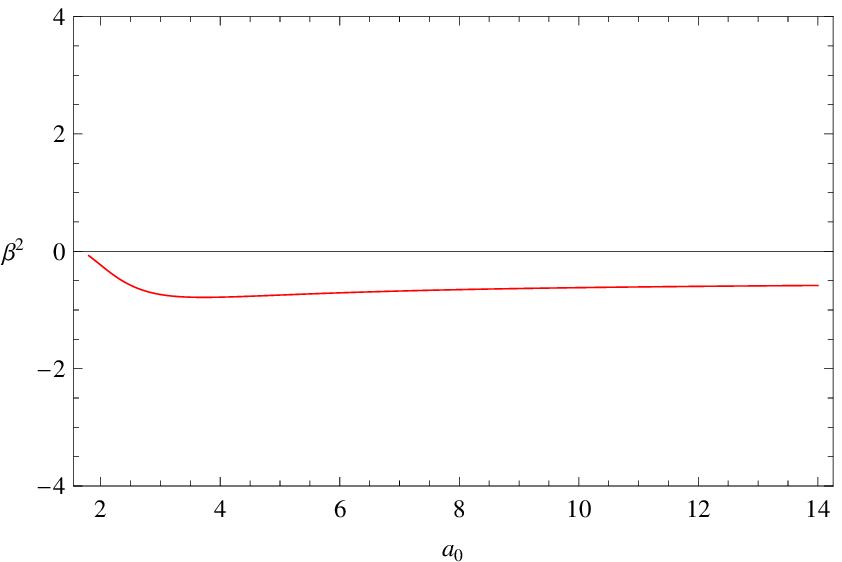,width=.30\linewidth,
height=2in}\caption{(Color online) Stability regions for thin-shell
wormholes corresponding to $Q=0.1,~0.5,~0.9$ and fixed value of
$M=1$.}
\end{figure}
\begin{eqnarray}\label{30}
{\sigma}'(a)+2{p}'(a)={\sigma}'(a)\left[1+2\frac{{p}'(a)}{{\sigma}'(a)}\right]
={\sigma}'(a)(1+2{\beta}^2).
\end{eqnarray}
Using the above equation and taking derivative of Eq.(\ref{28}), it
follows that
\begin{equation}\label{33}
V''(a)=f''(a)-8{\pi}^2(\sigma+2p)^2-8{\pi}^2
\left[2\sigma(1+2{\beta}^2)(\sigma+p)\right].
\end{equation}
For the stable configuration, we require that $V(a_0)=0=V'(a_0)$ and
$V''(a_0)>0$. It is easy to see that using Eq.(\ref{8}) in
(\ref{9}), the potential and its derivative are zero at $a=a_0$.
Solving the above equation for $\beta^2$ by letting $V''(a_0)=0$, we
have
\begin{equation}\label{34}
{\beta}^2=-\frac{1}{2}+\frac{\frac{f''}{8{\pi}^2}-(\sigma+2p)^2}{4\sigma(\sigma+p)},
\end{equation}

The graphical representation (Fig. \textbf{9}) shows the stability
regions to the left and right of the asymptote. To the right of
asymptote, the stability criterion $V''(a_0)>0$ leads to the
stability region with inequality
\begin{equation}\label{35}
{\beta}^2<-\frac{1}{2}+\frac{\frac{f''}{8{\pi}^2}-(\sigma+2p)^2}{4\sigma(\sigma+p)}.
\end{equation}
However, the sense of inequality for the stability region to the
left of the asymptote is reversed as
\begin{equation}\label{36}
{\beta}^2>-\frac{1}{2}+\frac{\frac{f''}{8{\pi}^2}-(\sigma+2p)^2}{4\sigma(\sigma+p)}.
\end{equation}
The regions of stability corresponding to $Q=0.1,~0.5,~0.9$ for
fixed value of $M=1$ are shown in Fig. \textbf{9}. It is noted that
the stability regions exist under the radial perturbations but do
not correspond to the range $0<\beta^2\leq{1}$ for different values
of charge.

\section{Thin-Shell Wormhole With $\Lambda$}

Here, we construct thin-shell wormhole from regular black hole with
cosmological constant and find its stability regions for some fixed
values of $Q$, $M$ and $\Lambda$. The event horizons for such a
black hole are shown in Fig. \textbf{10}. In this scenario, the
surface energy density and surface pressure are
\begin{eqnarray}\label{37}
\sigma&=&-\frac{1}{2\pi{a}}\left[1-\frac{2Ma^2}{(a^2+Q^2)^{\frac{3}{2}}}
+\frac{Q^2a^2}{(a^2+Q^2)^2}-\frac{\Lambda{a^2}}{3}\right]^\frac{1}{2},\\\label{38}
p&=&\frac{\left[1-\frac{4Ma^2}{(a^2+Q^2)^{\frac{3}{2}}}
+\frac{3Ma^4}{(a^2+Q^2)^{\frac{5}{2}}}
+\frac{2Q^2a^2}{(a^2+Q^2)^2}-\frac{2Q^2a^4}{(a^2+Q^2)^3}-\frac{\Lambda{a^2}}{3}\right]}
{4\pi{a}\left[1-\frac{2Mr^2}{(r^2+Q^2)^{\frac{3}{2}}}
+\frac{Q^2r^2}{(r^2+Q^2)^2}-\frac{\Lambda{a^2}}{3}\right]^\frac{1}{2}}.
\end{eqnarray}
Using Eq.(\ref{12}), the EoS parameter is given as
\begin{figure}
\centering \epsfig{file=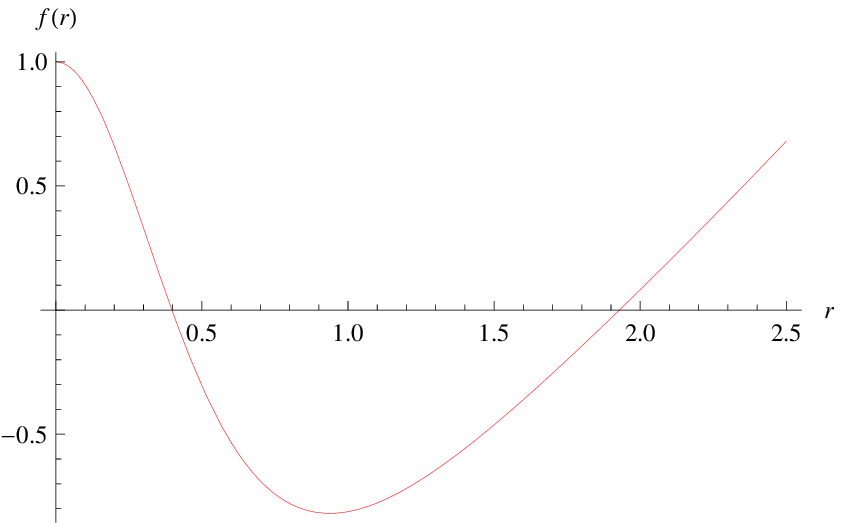,width=.48\linewidth,
height=1.5in} \epsfig{file=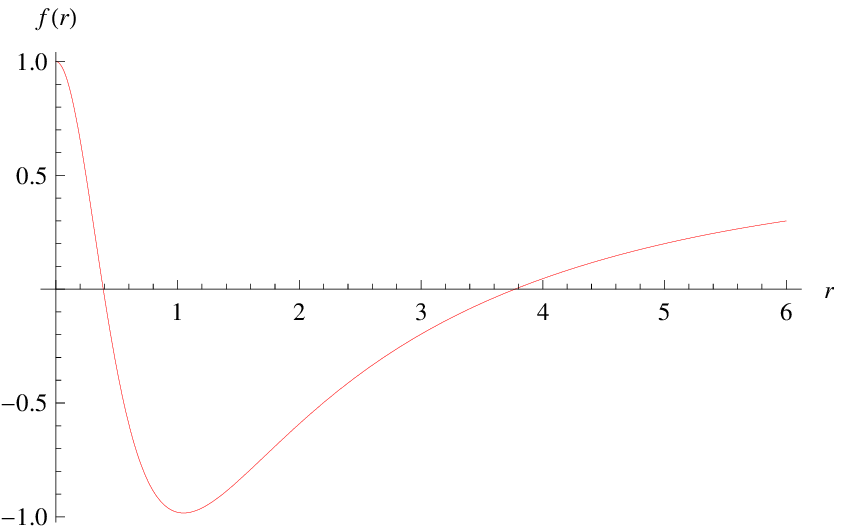,width=.48\linewidth,
height=1.5in} \caption{(Color online) Both left and right graphs
show event horizons ($f(r)$ cuts the $r-$axis) for fixed values of
$Q=0.7$ and $M=2$ corresponding to positive and negative $\Lambda$.}
\end{figure}
\begin{eqnarray}\label{39}
\omega=\frac{p}{\sigma}=-\frac{1}{2}-\frac{\left[-\frac{2Ma^2}{(a^2+Q^2)^{\frac{3}{2}}}
+\frac{3Ma^4}{(a^2+Q^2)^{\frac{5}{2}}}
+\frac{Q^2a^2}{(a^2+Q^2)^2}-\frac{2Q^2a^4}{(a^2+Q^2)^3}-\frac{\Lambda{a^2}}{3}\right]}
{2\left[1-\frac{2Ma^2}{(a^2+Q^2)^{\frac{3}{2}}}
+\frac{Q^2a^2}{(a^2+Q^2)^2}-\frac{\Lambda{a^2}}{3}\right]}.
\end{eqnarray}
The matter distribution in the shell is of dark energy type similar
to the above case, i.e., for $a\rightarrow\infty,$ we have
$\omega\rightarrow{-1}$.

The corresponding gravitational field will be
\begin{eqnarray}\label{40}
a^r=\left[-\frac{2Mr}{(r^2+Q^2)^{\frac{3}{2}}}
+\frac{3Mr^3}{(r^2+Q^2)^{\frac{5}{2}}}
+\frac{Q^2{r}}{(r^2+Q^2)^2}-\frac{2Q^2r^3}{(r^2+Q^2)^3}-\frac{\Lambda{r}}{3}\right].
\end{eqnarray}
The attractive $(a^r>0)$ and repulsive $(a^r<0)$ behavior of the
thin-shell wormhole with cosmological constant is shown in Fig.
\textbf{11} for fixed values of $M=1,~Q=0.1,~\Lambda=0.5$ and
$M=0.5,~Q=0.1,~\Lambda=-0.5$.

The total amount of exotic matter around the throat turns out to be
\begin{equation}\label{41}
\Omega_\alpha=-2a\left[1-\frac{2Ma^2}{(a^2+Q^2)^{\frac{3}{2}}}
+\frac{Q^2a^2}{(a^2+Q^2)^2}-\frac{\Lambda{a^2}}{3}\right]^{\frac{1}{2}}.
\end{equation}
Similar to the above case, the total amount of exotic matter can be
reduced by choosing some appropriate fixed values of $M,~Q$ and
$\Lambda$.
\begin{figure}
\centering \epsfig{file=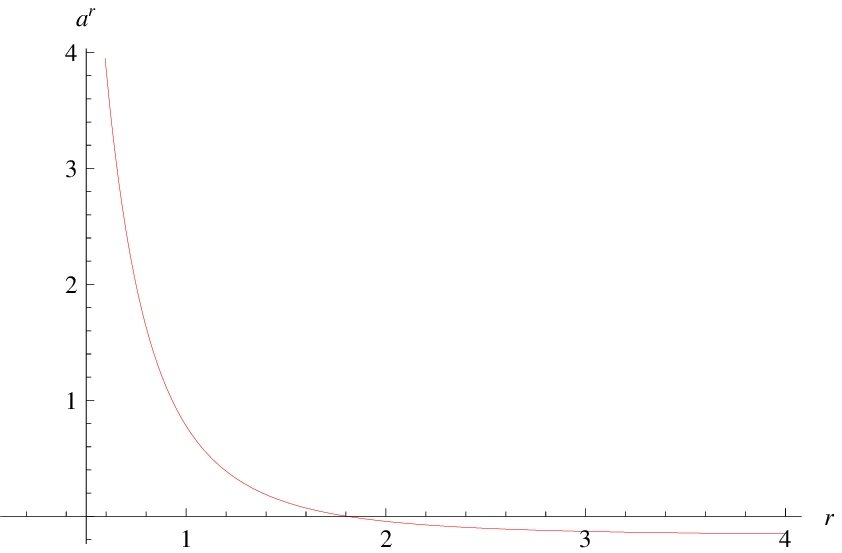,width=.48\linewidth,
height=1.5in}\epsfig{file=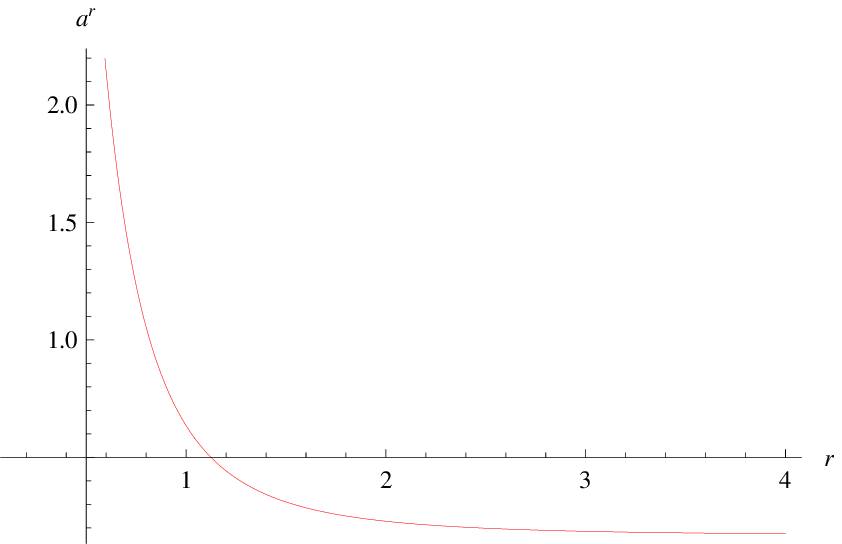,width=.48\linewidth,
height=1.5in} \caption{(Color online) Plot of $a^r$ versus radial
coordinate for positive and negative cosmological constant.}
\end{figure}

\subsection*{Stability Analysis}

Using the same procedure as in the absence of cosmological constant,
we have stability condition for the thin-shell wormhole in the
presence of cosmological constant as
\begin{equation}\label{42}
{\beta}^2=-\frac{1}{2}+\frac{\frac{f''}{8{\pi}^2}-(\sigma+2p)^2}{4\sigma(\sigma+p)},
\end{equation}
\begin{figure}
\centering \epsfig{file=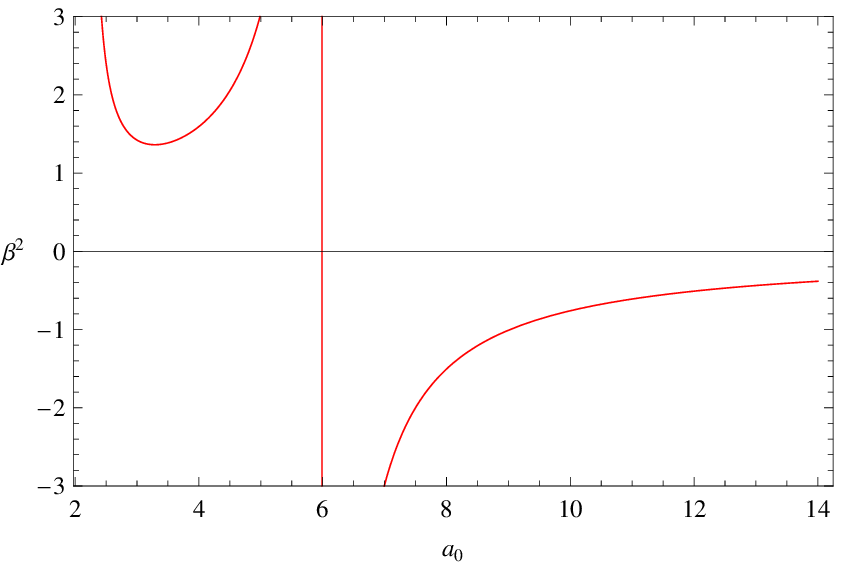,width=.32\linewidth,
height=1.8in} \epsfig{file=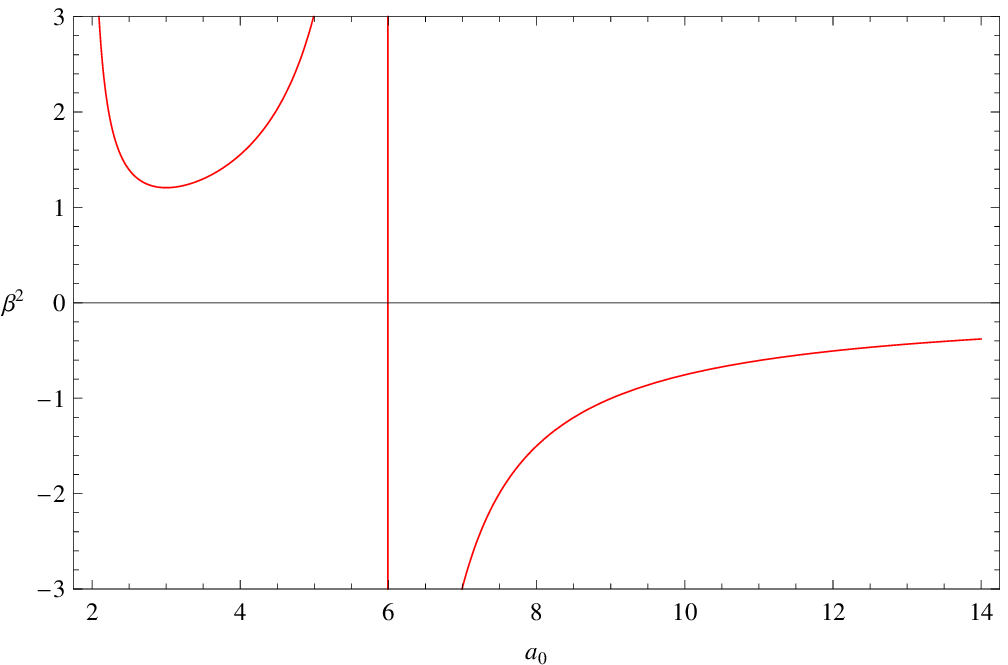,width=.32\linewidth,
height=1.8in} \caption{(Color online) Stability regions for
thin-shell wormholes corresponding to $\Lambda=-0.5,~-0.9$ for fixed
values of $Q=0.1$ and $M=2$.}
\end{figure}
\begin{figure}
\centering \epsfig{file=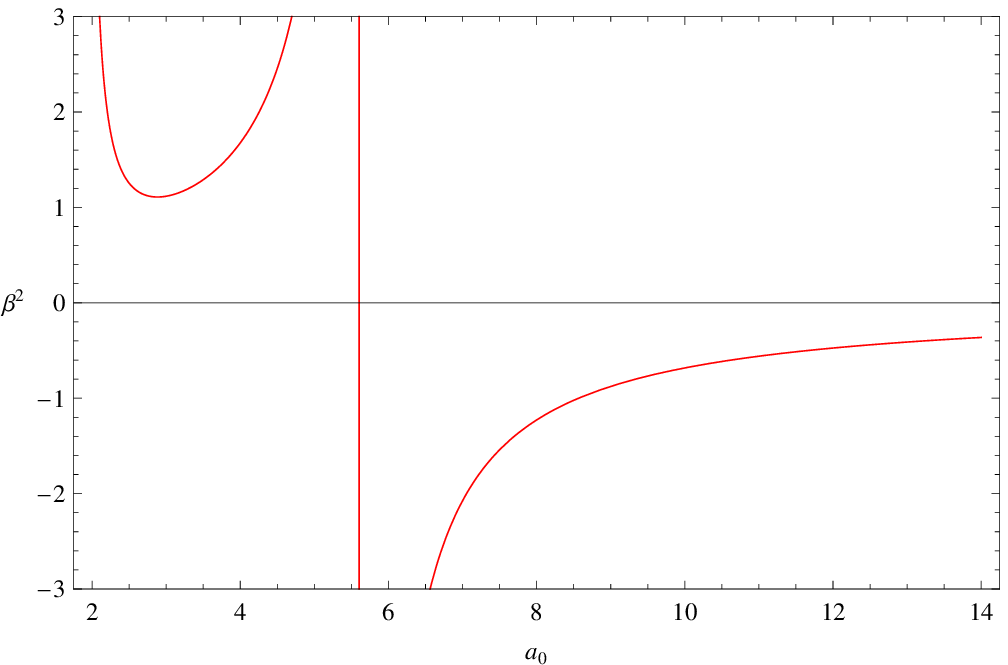,width=.32\linewidth,
height=1.8in} \epsfig{file=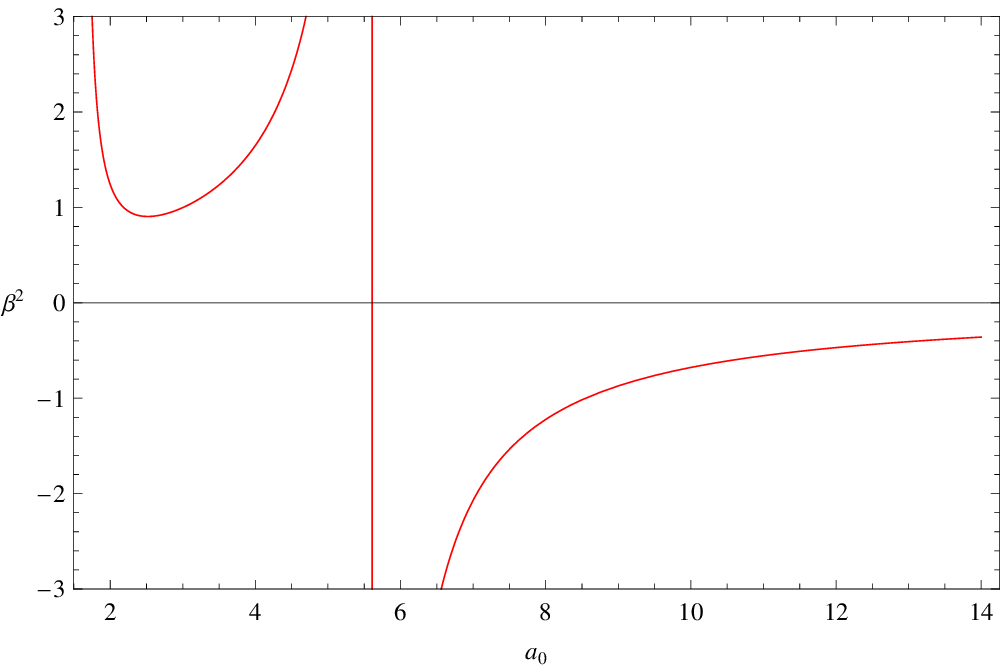,width=.32\linewidth,
height=1.8in}\caption{(Color online) Stability regions for
thin-shell wormholes corresponding to $\Lambda=-0.5,~-0.9$ for fixed
values of $Q=0.7$ and $M=2$}
\end{figure}
\begin{figure}
\centering \epsfig{file=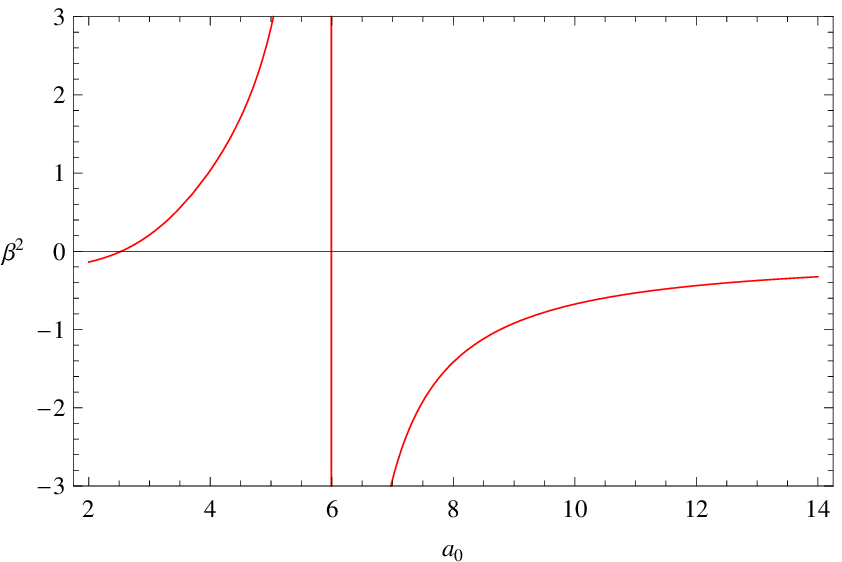,width=.32\linewidth,
height=1.8in} \epsfig{file=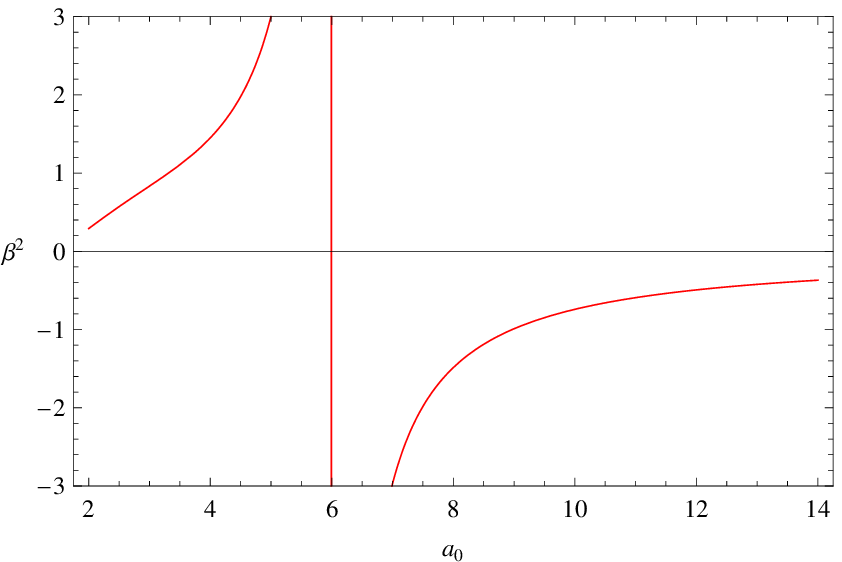,width=.32\linewidth,
height=1.8in}\caption{(Color online) Stability regions for
thin-shell wormholes correspond to $\Lambda=0.1,~0.9$ for fixed
values of $Q=0.1$ and $M=2$}
\end{figure}
where surface energy density and surface potential are given by
Eqs.(\ref{37}) and (\ref{38}). The typical regions of stability are
shown in Fig. \textbf{12} for fixed values of the parameters
$M=2,~Q=0.1,~\Lambda=-0.5$ and $M=2,~Q=0.1,~\Lambda=-0.9$. We see
that the stability regions can be extended by reducing the value of
cosmological constant. Further, we obtain the stability regions by
increasing the value of charge and keeping the same value of mass
and cosmological constant, i.e., $M=2,~Q=0.7,~\Lambda=-0.5$ and
$M=2,~Q=0.7,~\Lambda=-0.9$. It is observed that the stability
regions increase with the increase of charge which extends to the
region $0<\beta^2\leq1$, as shown in Fig. \textbf{13}.

On the other hand, we plot stability regions for positive
cosmological constant for fixed value of mass and charge. Figure
\textbf{14} shows the stability regions for
$M=2,~Q=0.1,~\Lambda=0.1$ and $M=2,~Q=0.1,~\Lambda=0.9$. We see from
Fig. \textbf{15} that the stability regions for
$M=2,~Q=0.7,~\Lambda=0.1$ and $M=2,~Q=0.7,~\Lambda=0.9$ are not of
typical form like in the case of negative cosmological constant.
\begin{figure}
\centering \epsfig{file=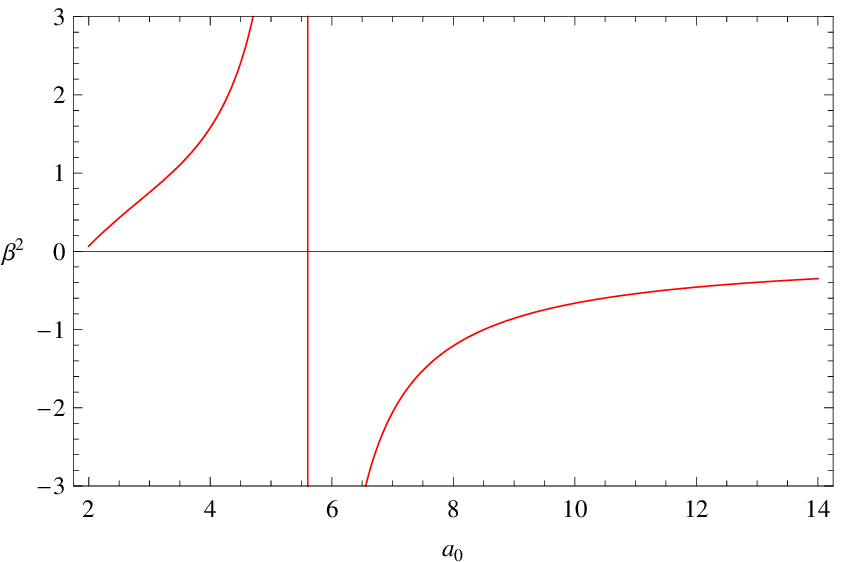,width=.32\linewidth,
height=1.8in} \epsfig{file=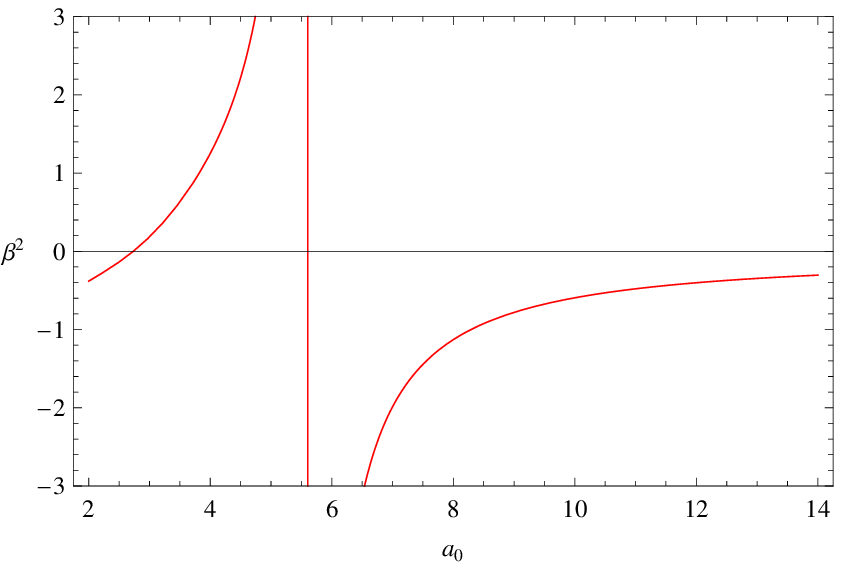,width=.32\linewidth,
height=1.8in}\caption{(Color online) Stability regions for
thin-shell wormholes correspond to $\Lambda=0.1,~0.9$ for fixed
values of $Q=0.7$ and $M=2$}
\end{figure}

\section{Summary}

We have studied various aspects of thin-shell wormholes constructed
from regular charged black hole with and without cosmological
constant in a nonlinear electrodynamics field. We have plotted
$\sigma$ and $p$ for different values of parameters $M$, $Q$ and
$a=a_0$ to show the presence of exotic matter confined within the
shell of the wormhole. It is observed that the matter distribution
is of the phantom energy type in both cases with and without
$\Lambda$. The nature of the wormhole (attractive and repulsive) has
been investigated for fixed values of parameters $M$ and $Q$. The
amount of exotic matter required to support the wormhole is always a
crucial issue. We have shown the variation of exotic matter
graphically with respect to charge and mass. It is found that the
total amount of exotic matter can be reduced by either decreasing
the value of charge or increasing the value of mass.

We have also addressed the issue of stability of thin-shell
wormholes subject to linearized radial perturbation around a static
solution. For this purpose, the stability regions have been plotted
in the form of parameter $\beta$. Figures \textbf{12-15} show the
stability regions to the left and right of the vertical asymptotes
for fixed values of $M,~Q,$ and $\Lambda$. Eiroa$^{{56})}$ has found
that the dilaton thin-shell wormholes have stable configurations in
small range as compared to Reissner-Nordstr$\ddot{o}$m wormholes for
the same value of charge. But in both cases, the stable
configurations do not lie in the realistic range i.e.,
$0<\beta^2\leqslant{1}$. However, Rahaman et al.$^{{16})}$ and
Usmani et al.$^{{17})}$ found the stable configurations in this
range. In our case, we have found that the stability regions belong
to this range for the case of negative cosmological constant by
varying its range for fixed values of charge and mass. This shows
the physical relevance of negative cosmological constant on wormhole
stability. In other cases, these stable configurations exist similar
to Ref.$^{{56})}$ but do not correspond to the range
$0<\beta^2\leqslant{1}$ for different values of the parameters.

\vspace{0.5cm}

{\bf Acknowledgments}

\vspace{0.5cm}

We would like to thank the Higher Education Commission, Islamabad,
Pakistan, for its financial support through the {\it Indigenous
Ph.D. 5000 Fellowship Program Batch-VII}. One of us (MA) would like
to thank University of Education, Lahore for the study leave.\\\\
1) N. S. Kardashev: Int. J. Mod. Phys. D \textbf{16} (2007) 909.\\
2) E. A. Larranaga: arXiv:gr-qc/0505054v6.\\
3) M. Safonova and D. F. Torres: Mod. Phys. Lett. A \textbf{17} (2002) 1685.\\
4) M. S. Morris and K. S. Thorne: Am. J. Phys. \textbf{56} (1988) 395.\\
5) M. Visser, S. Kar and N. Dadhich: Phys. Rev. Lett. \textbf{90} (2003) 201102.\\
6) M. Visser: Phys. Rev. D \textbf{39} (1989) 3182.\\
7) M. Visser: Nucl. Phys. B \textbf{328} (1989) 203.\\
8) E. Poisson and M. Visser: Phys. Rev. D \textbf{52} (1995) 7318.\\
9) M. Ishak and K. Lake: Phys. Rev. D \textbf{65} (2002) 044011.\\
10) C. Armendariz-Picon: Phys. Rev. D \textbf{65} (2002) 104010.\\
11) H. Shinkai and S. A. Hayward: Phys. Rev. D \textbf{66} (2002) 044005.\\
12) E. F. Eiroa and G. E. Romero: Gen. Relativ. Gravit. \textbf{36} (2004) 651.\\
13) F. S. N. Lobo and P. Crawford: Class. Quantum Grav. \textbf{21} (2004) 391.\\
14) M. Thibeault, C. Simeone and E. F. Eiroa: Gen. Relativ. Gravit. \textbf{38} (2006) 1593.\\
15) E. F. Eiroa: Phys. Rev. D \textbf{80} (2009) 044033.\\
16) F. Rahaman, Sk. A. Rahman, A. Rakib and Peter K. F. Kuhfitting:
Int. J. Theor. Phys. \textbf{49} (2010) 2364.\\
17) A. A. Usmani,  Z. Hasan, F. Rahaman, Sk. A. Rakib, R. Saibal and
Peter K. F. Kuhfitting: Gen. Relativ. Gravit. \textbf{42} (2010) 2901.\\
18) M. Sharif and M. Azam: JCAP \textbf{02} (2012) 043; Gen.
Relativ. Gravit. \textbf{44} (2012) 1181; Eur. Phys. J. C \textbf{73} (2013) 2407.\\
19) M. Visser and D. Hochberg: arXiv gr-qc/9710001.\\
20) S. A. Hayward: arXiv gr-qc/0203051.\\
21) A. Chodos and S. Detweiler: Gen. Relativ. Gravit. \textbf{14} (1982) 879.\\
22) G. Clement: Gen. Relativ. Gravit. \textbf{16} (1984) 131.\\
23) K. K. Nandi, B. Bhattacharjee, S. M. K. Alam and J. Evans: Phys. Rev. D \textbf{57} (1998) 823.\\
24) Y. G. Shen, H. Y. Guo, Z. Q. Tan and H. G. Ding: Phys. Rev. D \textbf{44} (1991) 1330.\\
25) S. Kar: Phys. Rev. D \textbf{53} (1996) 722.\\
26) L. A. Anchordoqui  and S. E. Perez Bergliaffa: Phys. Rev. D \textbf{62} (2000) 067502.\\
27) M. Born: \emph{On the Quantum Theory of the Electromagnetic
Field}, Proc. Roy. Soc. Lond. A \textbf{143} (1934) 410; M. Born and
L. Infeld, \emph{Foundations of the New Field Theory}, Proc. Roy.
Soc. A \textbf{144} (1934) 425.\\
28) J. F. Plebanski: \emph{Lectures on Non-linear Electrodynamics},
monograph of the Niels Bohr Institute Nordita, Copenhagen (1968).\\
29) E. A. Beato and A. Garcia: Phys. Rev. Lett. \textbf{80} (1998) 5056.\\
30) E. A. Beato and A. Garcia: Phys. Lett. B \textbf{464} (1999) 25;
Gen. Relativ. Gravit. \textbf{31} (1999)629.\\
31) N. Seiberg and E. Witten: JHEP \textbf{09} (1999) 032.\\
32) M. Novello, S. E. P. Bergliaffa and J. Salim: Phys. Rev. D
\textbf{69} (2004)127301.\\
33) P. V. Moniz: Phys. Rev. D \textbf{66} (2002) 103501.\\
34) R. Garcia-Salcedo and N. Breton: Class. Quantum Grav.
\textbf{20} (2003) 5425.\\
35) D. N. Vollick: Gen. Relativ. Gravit. \textbf{35} (2003) 1151.\\
36) K. A. Bronnikov: Phys. Rev. D \textbf{63} (2001)044005.\\
37) I. Dymnikova: Class. Quantum Grav. \textbf{21} (2004)4417.\\
38) N. Breton: Phys. Rev. D \textbf{72} (2005) 044015.\\
39) F. Baldovin, M. Novello, S. E. Perez Bergliaffa and J. M. Salim:
Class. Quant. Grav. \textbf{17} (2000) 3265.\\
40) A. Borde: Phys. Rev. D \textbf{50} (1994) 3392.\\
41) C. Barrabes and V. P. Frolov: Phys. Rev. D \textbf{53} (1996) 3215.\\
42) M. Mars, M. M. Martin-Prats and J. M. M. Senovilla: Class.
Quantum Grav. \textbf{13} (1996) L51.\\
43) A. Cabo and E. Ayon-Beato: Int. J. Mod. Phys. A \textbf{14} (1999) 2013.\\
44) S. W. Hawking and G. F. R. Ellis: \emph{The Large Scale
Structure of Spacetime}, (Cambridge University Press, 1975).\\
45) W. J. MO, R. G. Cai and R. K. SU: Commun. Theor. Phys. \textbf{46} (2006) 453.\\
46) H. Salazar, A. Garcia and J. Plebanski: J. Math. Phys.
\textbf{28} (1987) 2171.\\
47) D. Hochberg and M. Visser: Phys. Rev. D \textbf{56} (1997) 4745.\\
48) G. Darmois: \emph{Memorial des Sciences Mathematiques}
(Gautheir-Villars, 1927) Fasc. 25; W. Israel: Nuovo Cimento B \textbf{44} (1966) 1.\\
49) P. Musgrave and K. Lake: Class. Quantum Grav. \textbf{13} (1996) 1885.\\
50) M. Visser: \textit{Lorentzian Wormholes} (AIP Press, New York, 1996).\\
51) D. Hochberg and M. Visser: Phys. Rev. Lett.
\textbf{81} (1998) 746; Phys. Rev. D \textbf{58} (1998) 044021.\\
52) E. F. Eiroa and C. Simeone: Phys. Rev. D \textbf{71} (2005) 127501.\\
53) T. A. Roman: Phys. Rev. D \textbf{53} (1993) 5496.\\
54) L. Herrera: Phys. Lett. A \textbf{165} (1992) 206.\\
55) H. Abreu, H. Hernandez and L. A. Nunez: Class. Quantum Grav. \textbf{24} (2007) 4631.\\
56) E. F. Eiroa: Phys. Rev. D \textbf{78} (2008) 024018.\\
\end{document}